\documentclass[prd,twocolumn,superscriptaddress,floatfix,amsmath,amssymb,amsfonts]{revtex4-1}

\usepackage{float} 

\usepackage{comment}
\usepackage[normalem]{ulem}
\usepackage[english]{babel}
\usepackage{graphicx}
\usepackage{dcolumn}
\usepackage{bm}
\usepackage{blindtext}
\usepackage{verbatim}
\usepackage{mathrsfs}
\usepackage{tensor}
\usepackage{amsmath}
\usepackage{cancel}
\usepackage{physics}
\usepackage{epstopdf}
\usepackage{mathtools}
\usepackage{color}

\newcommand{\ii}{\mathrm{i}}

\usepackage[usenames,dvipsnames]{pstricks}
\usepackage{epsfig}
\usepackage{pst-grad} 
\usepackage{pst-plot} 

\usepackage{hyperref}

\usepackage{verbatim}

\begin{document}

\title{Particle detectors as witnesses for quantum gravity}

\author{R\'emi Faure}
\email{rfaure@perimeterinstitute.ca}
\affiliation{Perimeter Institute for Theoretical Physics, Waterloo, Ontario, N2L 2Y5, Canada}
\affiliation{Univ Lyon, Ens de Lyon, Univ Claude Bernard Lyon 1,
CNRS, Laboratoire de Physique, F-69342 Lyon, France}

\author{T. Rick Perche}
\email{trickperche@perimeterinstitute.ca}
\affiliation{Perimeter Institute for Theoretical Physics, Waterloo, Ontario, N2L 2Y5, Canada}

\affiliation{Instituto de F\'{i}sica Te\'{o}rica, Universidade Estadual Paulista, S\~{a}o Paulo, S\~{a}o Paulo, 01140-070, Brazil}

\author{Bruno de S. L. Torres}
\email{bdesouzaleaotorres@perimeterinstitute.ca}
\affiliation{Perimeter Institute for Theoretical Physics, Waterloo, Ontario, N2L 2Y5, Canada}

\affiliation{Instituto de F\'{i}sica Te\'{o}rica, Universidade Estadual Paulista, S\~{a}o Paulo, S\~{a}o Paulo, 01140-070, Brazil}

\begin{abstract}

We present a model for the coupling of non-relativistic quantum systems with a linearized gravitational field from a Lagrangian formulation. The coupling strongly resembles the light-matter interaction models that are known to be well approximated by the Unruh-DeWitt detector model for interactions with quantum fields. We then apply our model to linearized quantum gravity, which allows us to propose a detector based setup that can in principle probe the quantum nature of the gravitational field.


\end{abstract}

\maketitle

\section{Introduction}    

    One of the greatest challenges of modern theoretical physics is a consistent theory of quantum gravity. While there are numerous proposals (most notably, string theory and loop quantum gravity), the lack of experimental data makes it hard to favour any one theory
    . There have also been important debates about whether we need a quantum theory for the gravitational field at all. One of the difficulties is the lack of consensus about an experiment that would be able -- even in principle -- to distinguish the classical/quantum nature of gravity. While the scales for which we could observe quantum phenomena for gravity are already an issue, it is also unclear how to describe the interaction of quantum systems with the gravitational field. 
    
    One of the proposals of thought experiments that investigate the possible quantum nature of gravity is the Bose-Marletto-Vedral (BMV) experiment \cite{BMV1, BMV2}. It is an example of an attempt to get insight into the behaviour of quantum systems when interacting via the gravitational field. 
    Although this experiment has been widely debated and criticized \cite{BeiLok,BeiLok2}, some issues were addressed in \cite{flaminia} and an explanation of the framework treating gravity as a quantum field has been given in \cite{Mazumdar}.
    
    If one is to approach the nature of gravity, one should be able to not only find gravitational effects that are generated by quantum sources, but also to probe effects that are intrinsically associated to properties of quantum fields. One of the ways of doing that is to look for an experimental setup that could be described either by treating gravity as a classical field or as a quantum one which yields different results in both cases.
    
    A known way of quantizing gravity in low energy regimes is by considering a small perturbation around a fixed background metric, such as the flat one \cite{weinberg}. The Einstein-Hilbert Lagrangian can then be linearized to second order to yield a quadratic Lagrangian for the perturbation. The equations of motion can then be solved analytically in Minkowski spacetime \cite{carroll}. Canonical quantization then provides a standard procedure for quantizing the perturbation by introducing commutation relations that give rise to creation and annihilation operators. Once this procedure is complete, this quantum field  can be associated to gravitational perturbations and its excitations are regarded as gravitons. 
    


    The issue of probing quantum fields is highly nontrivial. One way of approaching the problem is with the use of particle detector models, such as the Unruh-DeWitt (UDW) detector \cite{DeWitt,Unruh-Wald}. This model consists of a localized two-level system that couples to a local degree of freedom of the field. It can be shown to be a good approximation for the light-matter interaction, and has become ubiquitous in various topics in relativistic quantum information. Its uses range from quantum optics to quantum field theory in curved spacetimes. 
    
    One important feature of UDW detectors is to give operational means by which one could explore quantum aspects of a field. One of such aspects is the entanglement structure of the vacuum state in a quantum field theory. This can be realized by a process called \textit{entanglement harvesting}, which is based on studying the dynamics of a pair of spatially separated detectors that couple locally to a quantum field \cite{Valentini1991, Reznik1, Pozas-Kerstjens:2015}. It can be shown that the detectors can become entangled through an interaction with the field, 
    even if the two interaction regions are spacelike separated. 
    This effect has no classical analogue, and it provides a criterion by which one could identify a field as being genuinely quantum. Therefore, if one could detect entanglement generated by locally coupling detectors to gravity, that would constitute strong evidence for the quantum nature of the gravitational field. It has been shown \cite{robMann1,robMann2} that entanglement harvesting with quantum fields is also sensitive to the geometry and topology of the background spacetime, and therefore one can probe the classical effects of a curved background. We stress that although quantum field theory in curved spacetimes probes gravitational effects in quantum fields, it does not probe quantum effects on the gravitational field itself.

    This paper is organized as follows. Sections \ref{UDW}, \ref{lineargrav} and \ref{entharv} contain a review of the relevant aspects of the Unruh-DeWitt model, linearized quantum gravity on Minkowski background, and the entanglement harvesting protocol, with section \ref{sectionNoHarvest} emphasizing how entanglement harvesting is indeed only possible via quantum fields. Section \ref{sectionCoupling} shows how the coupling of non-relativistic quantum system with linearized perturbations of the gravitational field can be well modelled by an UDW-like detector. The coupling closely resembles the one of atoms to the electromagnetic field (i.e., the light-matter interaction). The idea is then to consider two such systems (e.g. atoms) and perform entanglement harvesting on them. If one could prevent the electromagnetic field from providing a source of entanglement for the two detectors, only the gravitational field would be capable of entangling them. This could be done, for instance, if they were put in separate optical cavities. Therefore, if the systems could be shown to be in an entangled state, the entanglement would have to come from interactions with the vacuum of the gravitational field. Section \ref{Discussions} provides a rough estimate on how relevant the coupling with the gravitational field would be in comparison with the electromagnetic coupling. This gives an order of magnitude to when one should expect to be able to see effects from the quantum nature of the gravitational field in this model.
    
    The present work not only proposes a new framework for probing the quantum nature of gravity, but it is also helps us gauge the difficulties in detecting quantum gravitational effects. It also provides a new way of thinking about the interaction of non-relativistic quantum systems with quantum fields. By thinking in terms of an effective Lagrangian description for the wavefunction of such systems, one can derive a form of coupling between detectors and fields that works similarly for the electromagnetic and gravitational cases. This allows us to extend the available techniques of UDW detectors, which are known to be applicable to a wide range of phenomena.

\section{Unruh-DeWitt Detectors}\label{UDW}
    
    An Unruh-DeWitt detector is a model of particle detectors introduced by Unruh and DeWitt in \cite{DeWitt,Unruh-Wald} that consists of a localized non relativistic quantum system that couples locally to a quantum field. This model has been shown to be a good approximation for the description of the light-matter interaction in the context of quantum optics, as shown in \cite{Pozas2016,eduardo}.
    
    A possible setup consists of a two-level system that couples to a real scalar quantum field in the vicinity of the trajectory of its center of mass, parametrized by a timelike trajectory ${\mathsf z(\tau) = (t(\tau),\bm{z}(\tau))}$, with proper time $\tau$. The free Hamiltonian for the detector generates the free evolution of the two-level system with respect to its proper time $\tau$, and is given by
    \begin{equation}\label{HD}
        \hat H_{d}^\tau = \Omega \hat \sigma^{+}\hat \sigma^{-},
    \end{equation}
    where $\Omega$ is the energy gap of the detector and the operators $\hat{\sigma}^{\pm}$ are the $SU(2)$ ladder operators. In the basis of eigenstates of $\hat{H}_d^\tau$, labeled as $\{\ket{e}, \ket{g}\}$ (excited and ground states, respectively), we have
    \begin{equation}
    \begin{gathered}
        \hat{\sigma}^+ = \ket{e}\bra{g}, \\
        \hat{\sigma}^- = \ket{g}\bra{e}.
        \end{gathered}
    \end{equation}
    
    The quantum field, on the other hand, is described in terms of a complete set of solutions to the Klein-Gordon equation, given by modes $u_{\bm{k}}(\mathsf{x})$. The free dynamics of the field can then be solved to give
    
    \begin{equation}\label{field}
     \phi(\mathsf x) = \int \dd^{n} \!\bm{k}\left(a^\dagger_{\bm{k}} u_{\bm{k}}(\mathsf x)+a_{\bm{k}} u^*_{\bm{k}}( \mathsf x) \right),
\end{equation}
where $a^\dagger_{\bm{k}}$ and $a_{\bm{k}}$ are creation and annihilation operators associated to the choice of modes given by $\{u_{\bm{k}} (\mathsf{x})\}$, and we write $\mathsf{x} = (t,\bm x)$ for the spacetime points in a given coordinate system.
    
The coupling between detector and field should then be able to promote excitations and de-excitations of the field, while also promoting transitions between the two levels of the detector. The simplest way to achieve this is by promoting a linear coupling between the field and the monopole operator of the detector. In the case where we consider the detector system to have some nontrivial spatial extension, the interaction Hamiltonian is given by
\begin{equation}\label{Hint}
     \hat H_I^\tau(\tau) = \lambda \chi(\tau)\hat{\mu}(\tau)\int_{\Sigma_\tau}\!\!\!\dd^n\bar{\bm x}\sqrt{\bar{g}}\: f(\bm{\bar{x}})  \hat \phi(\bar{\mathsf{x}}),
\end{equation}
where $\hat{\mu}(\tau) = e^{\ii\Omega \tau}\hat \sigma^+ +e^{-\ii\Omega\tau}\hat \sigma^-$ is the monopole operator of the detector, $\chi(\tau)$ is a switching function that determines the temporal profile of the coupling between detector and field, and $f(\bm{\bar{x}})$ is a smearing function which dictates the spatial extension of the interaction. $\Sigma_{\tau}$ are spacelike surfaces of constant $\tau$ (i.e., they are simultaneity surfaces associated to the proper time of the center of mass of the detector) and $\bm{\bar{x}}$ are spatial coordinates on such surfaces. Together, $(\tau, \bm{\bar{x}})$ correspond to Fermi normal coordinates associated to the trajectory of the detector's center of mass frame. For more details, see \cite{us}.

   This model for particle detectors is well motivated by realistic, first principle approaches to the light-matter interaction, and is very effective at describing transitions without exchange of angular momentum between detector and field. In the sections that follow, we will show how a similar argument can be applied to the interaction between an atom and an external gravitational field, which makes the Unruh-DeWitt approach to detecting gravitons a reasonable first approximation. 
   
       \section{Linearized Quantum Gravity}\label{lineargrav}
    
    The standard concept of perturbations in general relativity starts with a background metric that satisfies Einstein's equations and adds to it a rank two tensor, which is regarded as a perturbation. In the case that we are interested in, we will set the background metric to be Minkowski, denoted by $\eta_{\mu\nu}$ and add to it the perturbation $h_{\mu\nu}$, so that the full metric is given by
    \begin{equation}
        g_{\mu\nu} = \eta_{\mu\nu}+h_{\mu\nu}.
    \end{equation}
    
    By writing the Einstein Hilbert action to second order, it is possible to obtain a free Lagrangian for the perturbation $h_{\mu\nu}$ \cite{carroll}. After gauge fixing, it can be shown that the equations of motion for the field read
    \begin{equation}
        \Box\: {h}_{\mu\nu} = 0.
    \end{equation}
    This allows us to write the general solution in terms of the two independent polarization degrees of freedom, $\epsilon^{(1)}_{\mu\nu}(\bm k)$ and $\epsilon^{(2)}_{\mu\nu}(\bm k)$ and the free modes of momentum $\bm k$, $a_{\bm k}^{(\lambda)},a_{\bm k}^{(\lambda)*}$:
    \begin{equation}
        {h}_{\mu \nu}(\mathsf{x}) \!=\! \frac{1}{(2\pi)^{\frac{3}{2}}}\!\sum_{\lambda =1}^2\int\!\!\! \frac{\mathrm{d}^{3} \bm k}{\sqrt{2\abs{\bm k}}}\!\left({a}_{\bm k}^{(\lambda)} e^{\ii \mathsf{k}\cdot \mathsf{x}}\! +\!{a}_{\bm k}^{(\lambda)*} e^{- \ii \mathsf{k}\cdot \mathsf{x}}\! \right)\!\epsilon^{(\lambda)}_{\mu\nu}(\bm{k}).
    \end{equation}
    In the so called traceless transverse gauge, the polarization tensors satisfy $k^\mu\epsilon^{(\lambda)}_{\mu\nu}(\bm{k}) = 0$ and $\epsilon\indices{^\mu_\mu^{(\lambda)}}(\bm{k}) = 0$, which also implies that the vacuum solution for $h_{\mu\nu}(\mathsf{x})$ is traceless.
    
    We then quantize the free modes by promoting them to the creation and annihilation operators and imposing the canonical commutation relations,
    \begin{equation}
        \comm{\hat{a}_{\bm k}^{(\lambda)}}{\hat{a}_{\bm k'}^{(\lambda')\dagger}} = \delta^{\lambda\lambda'}\delta^{(3)}(\bm k-\bm k')\openone.
    \end{equation}
    We can then write the quantized gravitational perturbation as
    \begin{equation}\label{modeExpansionPerturbation}
        \hat{h}_{\mu \nu}(\mathsf{x}) \!=\! \frac{1}{(2\pi)^{\frac{3}{2}}}\!\sum_{\lambda =1}^2\int\!\!\! \frac{\mathrm{d}^{3} \bm k}{\sqrt{2\abs{\bm k}}}\!\left(\hat{a}_{\bm k}^{(\lambda)} e^{\ii \mathsf{k}\cdot \mathsf{x}}\! +\!\hat{a}_{\bm k}^{(\lambda)\dagger} e^{- \ii \mathsf{k}\cdot \mathsf{x}}\! \right)\!\epsilon^{(\lambda)}_{\mu\nu}(\bm{k}).
    \end{equation}
    The operator above has the exact same expression as the electromagnetic field in the Lorentz gauge, but with polarization tensors instead of vectors. Notice that for the above expression to be valid, we must be using Planck units, otherwise we would get a factor proportional to the Planck mass to ensure that the final solution for $\hat{h}_{\mu\nu}(\mathsf{x})$ is dimensionless.
   
\section{Entanglement Harvesting}\label{entharv}
    
    One of the most remarkable facts about quantum field theory is that the vacuum state contains correlations between both spacelike and timelike separated regions. This feature is not only interesting from the fundamental perspective, but it is also useful for many applications on the area of relativistic quantum information, such as the Unruh-Hawking effect \cite{Unruh1976, Sciama1977}, quantum energy teleportation \cite{teleportation} and entanglement  harvesting \cite{Pozas-Kerstjens:2015}. 
    
    Introduced by Valentini \cite{Valentini1991}, later explored by Reznik \cite{Reznik1,reznik2}, and generalized for extended detectors by Mart\'in-Martinez \textit{et al}. \cite{Pozas-Kerstjens:2015}, the technique of entanglement harvesting consists on the coupling of two Unruh-DeWitt detectors to a quantum field, but not with each other. Even though they are not directly coupled, it can be shown that after time evolution the final state of the detectors can be an entangled state. In the next subsection we briefly review the results obtained for harvesting in flat spacetimes with the use of smeared detectors.

    \subsection{Entanglement from the quantum vacuum}
    
    Here we will specialize to the case of inertial detectors in flat spacetime, coupled to a massless real scalar field, as was done in \cite{Pozas-Kerstjens:2015}. In this scenario, the coordinates $\mathsf{x} = (t,\bm x)$ of equation \eqref{field} reduce to the standard Cartesian coordinates in Minkowski spacetime. Let us label the detectors $A$ and $B$ with respective energy gaps $\Omega_A$ and $\Omega_B$, such that their free Hamiltonians will be given, as in equation \eqref{HD}, by
    \begin{align}
        \hat{H}_{A} &= \Omega_A \hat{\sigma}^+_{(A)}\hat{\sigma}^-_{(A)},\label{HA}\\
        \hat{H}_{B} &= \Omega_B \hat{\sigma}^+_{(B)}\hat{\sigma}^-_{(B)}. \label{HB}
    \end{align}
    In this setting, the interaction Hamiltonian of two UDW detectors coupled to the scalar field becomes the sum of the interaction Hamiltonians for each of the detectors,
    \begin{equation}\label{HII}
        \hat{H}_I(t) = \hat{H}_{I,A}(t)+\hat{H}_{I,B}(t).
    \end{equation}
    Here $\hat{H}_{I,A}(t)$ and $\hat{H}_{I,B}(t)$ are the interaction Hamiltonians for the corresponding detectors $A$ and $B$. Equation \eqref{Hint} gives us, for inertial detectors at rest with respect to each other in flat spacetimes,
    \begin{align}
        \hat{H}_{I,A}(t) &= \lambda_{A} \chi_{A}(t) \hat{\mu}_{A}(t) \int \mathrm{d}^{n} \bm x f_{A}(\bm x) \hat{\phi}(\mathsf{x}),\\
        \hat{H}_{I,B}(t) &= \lambda_{B} \chi_{B}(t) \hat{\mu}_{B}(t) \int \mathrm{d}^{n} \bm x f_{B}(\bm x) \hat{\phi}(\mathsf{x}).
    \end{align}
    
    The time evolution operator in the interaction picture can then be expressed as a Dyson expansion in terms of the interaction Hamiltonian:    
    \begin{equation}
    \begin{gathered}
        \hat{U}
        =\openone \underbrace{-\mathrm{i} \int_{-\infty}^{\infty}\!\!\!\! \mathrm{d} t \hat{H}_I(t)}_{\hat{U}^{(1)}} \underbrace{-\int_{-\infty}^{\infty} \!\!\!\!\mathrm{d} t \int_{-\infty}^{t}\!\!\!\! \mathrm{d} t^{\prime} \hat{H}_I(t) \hat{H}_I\left(t^{\prime}\right)}_{\hat{U}^{(2)}}+\mathcal{O}(\lambda^3),
    \end{gathered}
    \end{equation}
    where we use $\mathcal{O}(\lambda^3)$ to denote third order in products of the couplings $\lambda_A$ and $\lambda_B$.
    
    Now consider an initial state that is completely separable for the two detectors and the field. Moreover, we will consider that the initial state given by the field is in its vacuum state, $\ket{0}$, and both detectors in their respective ground states, such that
    
    \begin{equation}\label{initialstates}
    \begin{gathered}
        \hat{\rho}_{0}=|0\rangle\langle 0| \otimes \hat{\rho}_{A B, 0}, \\
        \hat{\rho}_{A B, 0}=\left|g_{A}\right\rangle\left\langle g_{A}|\otimes| g_{B}\right\rangle\left\langle g_{B}\right|.
        \end{gathered}
    \end{equation}
    The partial state of the detectors after their interaction, tracing out the states of the field, will then be given by
    
    \begin{equation}
        \hat{\rho}_{A B}=\operatorname{Tr}_{\phi}\left(\hat{U} \hat{\rho}_{0} \hat{U}^{\dagger}\right).
    \end{equation}
    In the basis 
    
    \begin{equation}
        \left\{\left|g_{A}\right\rangle \otimes\left|g_{B}\right\rangle,\left|e_{A}\right\rangle \otimes\left|g_{B}\right\rangle,\left|g_{A}\right\rangle \otimes\left|e_{B}\right\rangle,\left|e_{A}\right\rangle \otimes\left|e_{B}\right\rangle\right\},
    \end{equation}
    the reduced state can be shown to have the following matrix representation:
    \begin{equation} 
        \hat{\rho}_{\mathrm{AB}}=\left(\begin{array}{cccc}
1-\mathcal{L}_{A A}-\mathcal{L}_{B B} & 0 & 0 & \mathcal{M}^{*} \\
0 & \mathcal{L}_{A A} & \mathcal{L}_{A B} & 0 \\
0 & \mathcal{L}_{B A} & \mathcal{L}_{B B} & 0 \\
\mathcal{M} & 0 & 0 & 0
\end{array}\right)+\mathcal{O}\left(\lambda^{4}\right),
    \end{equation}
    where here we have defined, for $I,J = A,B$,
    \begin{equation}
        \begin{aligned}
\mathcal{L}_{I J} &=\int \mathrm{d}^{n} \boldsymbol{k} L_{I}(\boldsymbol{k}) L_{J}(\boldsymbol{k})^{*}, \\
\mathcal{M} &=\int \mathrm{d}^{n} \boldsymbol{k} M(\boldsymbol{k})
\end{aligned}
    \end{equation}
    and finally, $L_{I}(\bm{k})$ and $M(\bm{k})$ are given by
    \begin{equation}
        \begin{aligned}
            L_{I}(\boldsymbol{k})=& \lambda_{I} \frac{e^{-\mathrm{i} \boldsymbol{k} \cdot \boldsymbol{x}_{I}} \tilde{F}(\boldsymbol{k})}{\sqrt{2|\boldsymbol{k}|}} \int_{-\infty}^{\infty} \mathrm{d} t_{1} \chi_{I}\left(t_{1}\right) e^{\mathrm{i}\left(|\boldsymbol{k}|+\Omega_{I}\right) t_{1}}, \\
            M(\boldsymbol{k})=&-\lambda_{A} \lambda_{B} e^{\ii \boldsymbol{k} \cdot\left(\boldsymbol{x}_{A}-\boldsymbol{x}_{B}\right)} \frac{[\tilde{F}(\boldsymbol{k})]^{2}}{2|\boldsymbol{k}|} \\
            & \times \int_{-\infty}^{\infty} \mathrm{d} t_{1} \int_{-\infty}^{t_{1}} \mathrm{d} t_{2} e^{-\mathrm{i}|\boldsymbol{k}|\left(t_{1}-t_{2}\right)} \\
            &\left[\chi_{A}\left(t_{1}\right) \chi_{B}\left(t_{2}\right) e^{\mathrm{i}\left(\Omega_{A} t_{1}+\Omega_{B} t_{2}\right)}\right.\\
            &\left.+\chi_{B}\left(t_{1}\right) \chi_{A}\left(t_{2}\right) e^{\mathrm{i}\left(\Omega_{B} t_{1}+\Omega_{A} t_{2}\right)}\right].
            \end{aligned}
    \end{equation}
    The function $\tilde{F}(\bm{k})$ is given by
    \begin{equation}
        \tilde{F}(\boldsymbol{k})=\frac{1}{\sqrt{(2 \pi)^{n}}} \int \mathrm{d}^{n} \boldsymbol{x} F(\boldsymbol{x}) e^{i \boldsymbol{k} \cdot \boldsymbol{x}},
    \end{equation}
    and $F(\bm{x})$ is the function that dictates the spatial profile of both detectors. In other words, in order to get to the expressions above, we have also assumed similar smearing functions for the two detectors, given by
    
    \begin{equation}\label{smearing}
        \begin{gathered}
        f_A(\bm{x}) = F(\bm{x} - \bm{x}_A), \\
        f_B(\bm{x}) = F(\bm{x} - \bm{x}_B)
        \end{gathered}
    \end{equation}
    where $\bm{x}_A$ and $\bm{x}_B$ label the center of mass of detectors $A$ and $B$, respectively.

    Given the final reduced state of the two detectors, one might then ask about the resulting correlations and entanglement between them. For a two-qubit system, a simple measure of entanglement is given by the sum of the negative eigenvalues of the partially transposed density matrix (also called negativity). Up to second order in the coupling, there is only one eigenvalue for the partially transposed density matrix $\rho_{AB}$ that can possibly be negative, and that is given by
    \begin{equation}
        E_{1}\!=\!\frac{1}{2}\left[\mathcal{L}_{A A}\!+\!\mathcal{L}_{B B}\!-\!\sqrt{\left(\mathcal{L}_{A A}\!-\!\mathcal{L}_{B B}\right)^{2}\!+\!4|\mathcal{M}|^{2}}\right]+\mathcal{O}\left(\lambda^{4}\right),
    \end{equation}    
    which means that we can take as our measure of entanglement the negativity estimator $\mathcal{N} = \text{max}(0, -E_1)$. For more details on how the entanglement harvesting protocol described above depends on various characteristics of the setup (such as the size of the detectors' spatial extension, their energy gaps, the sudden vs. smooth types of switching, and the dimension of spacetime), see \cite{Pozas-Kerstjens:2015}.
    
    \newcommand{\mf}{\mathsf}
    \subsection{No entanglement from classical fields}\label{sectionNoHarvest}
    
    After presenting explicitly the procedure for harvesting entanglement from the vacuum state of quantum fields, we may wonder if the same effect would be possible when an UDW detector couples to a classical scalar field instead. In this case, the only quantum systems under consideration would be the two detectors themselves. These would still be modelled by two-level systems that evolve according to the free Hamiltonians \eqref{HA} and \eqref{HB} and undergo an inertial trajectory in Minkowski spacetime. We will model the interaction neglecting the backreaction of the detectors on the field. We do so taking under consideration the fact that if the detectors are spacelike separated, no influence one of them has on the field can propagate to the other one.
    
    The interaction Hamiltonian would then be prescribed in a similar way to \eqref{HII}, except for the fact that the field $\phi$ would be a classical function instead of a quantum operator. Assuming that the two detectors under consideration still follow inertial trajectories, the interaction Hamiltonian should be prescribed as
    
    \begin{align}\label{}
        \hat{H}^{c}_I(t) &=  \hat{H}^{c}_{I,A}(t) + \hat{H}^{c}_{I,B}(t),
    \end{align}
    where $\hat{H}^{c}_{I,A}(t)$ and $\hat{H}^{c}_{I,B}(t)$ are the interaction Hamiltonians for each of the detectors. These are explicitly given by:
    \begin{align}
        \hat{H}^{c}_{I,A}(t) = \lambda_A\chi_A(t) \hat{\mu}_{A}(t)\int \dd^n \bm{x}\: f_A({\bm x})\phi_c({\mathsf{x}}), \\
        \hat{H}^{c}_{I,B}(t) = \lambda_B\chi_B(t) \hat{\mu}_{B}(t)\int \dd^n \bm{x}\: f_B({\bm x})\phi_c({\mathsf{x}}),
    \end{align}
    where $f_A(\bm{x})$ and $f_B(\bm{x})$ are the smearing functions of the respective detectors, as given in equation \eqref{smearing}.
    
    The next step is then to calculate the time evolution operator, which is given by the time order exponential of the interaction Hamiltonian:
    \begin{align}
        \hat{U}^{c} = \mathcal{T} \exp \left( -\ii \int \dd t \: \hat{H}_I^{c}(t)\right).
    \end{align}
    Noting that the Hamiltonians $\hat{H}^c_{I,A}(t)$ and $\hat{H}^c_{I,B}(t)$ commute for every value of $t$, we can rewrite the above operator as a product of operators $\hat{U}^c_A$ and $\hat{U}^c_B$, defined by
    \begin{align}
        \hat{U}_A^c &= \mathcal{T} \exp\left(-\ii\int \dd t\: \hat{H}_{I,A}^{c}(t)\right),\\
        \hat{U}_B^c &= \mathcal{T} \exp\left(-\ii\int \dd t\: \hat{H}_{I,B}^{c}(t)\right).
    \end{align}
    In the end we get   
    $\hat{U}^c = \hat{U}_A^c\hat{U}_B^c$.
    
    If the detectors start in an unentangled state that factors into $\hat{\rho}_A$ and $\hat{\rho}_B$, the initial state of the full system of the two detectors can be described by a classical admixture of separable states, $\hat{\rho}_0 = \sum_{i} p_i\hat{\rho}^{(i)}_A\otimes\hat{\rho}^{(i)}_B$. Note that there is no quantum state corresponding to the field; the state of the field is identified with a classical configuration.
    
    After the interaction between the detectors and field takes place, we get the following final state for the detector
    \begin{equation}
        \hat{\rho}_{AB}  = \hat{U}^c\hat{\rho}_0 \hat{U}^{c\dagger} = \sum_{i}p_i\left(\hat{U}_A^c \hat{\rho}^{(i)}_A \hat{U}_A^{c\dagger}\right) \otimes \left(\hat{U}_B^c \hat{\rho}^{(i)}_B \hat{U}_B^{c\dagger}\right),
    \end{equation}
    which is again a state that may have classical correlations, but is not entangled. Therefore, no entanglement can be harvested by the interaction of Unruh-DeWitt detectors with a classical field, provided we neglect the back reaction of the detectors on the field. This approximation is well justified for spacelike separated regions, since in this case the perturbations caused by one of the detectors cannot influence the response of the second detector. Therefore, only quantum fields are able to produce entanglement, it cannot be produced by a classical field.

\section{The Coupling of Quantum Systems to Linearized Gravity}\label{sectionCoupling}
    
    
    In this section, we provide a model for the interaction of quantum systems undergoing inertial trajectories in flat spacetimes with weak gravitational fields (to linear order). This scenario can be described in a similar fashion to the one seen in light-matter interactions. The approach presented in \cite{Pozas2016} consists of starting with the Hamiltonian for an atom and introducing an external electromagnetic field that interacts with the system via minimal coupling. When restricting the dynamics to transitions between two energy levels, it can be shown that the interaction Hamiltonian is given by
    \begin{equation}\label{HIEM}
        \hat{H}_I\!=\!\!\! \int\!\! \dd^3 \bm x \!\Bigg(\!\!\bm F(\bm x)\cdot \hat{\bm E}(\mathsf{x}) e^{\ii\Omega t} \ket{e}\!\!\bra{g}+\bm F^*(\bm x)\cdot \hat{\bm E}(\mathsf{x}) e^{-\ii\Omega t}\ket{g}\!\!\bra{e}\!\!\!\Bigg).
    \end{equation}
    Here, $\hat{\bm E}(\mathsf{x}) = \hat{\bm E}(t,\bm x)$ is the free quantized electromagnetic field, $\ket{g},\ket{e}$ are the excited and ground states of the process  under consideration. $\Omega$ is the energy gap between the two levels and  $\bm F(\bm x)$ models the atom's dipole, which is given in terms of the eigenfunctions of the free Hamiltonian as
    \begin{equation} \label{fsmearingeg}
        \bm F(\bm x) =  \psi_e^*(\bm x) \bm x \psi_g(\bm x).
    \end{equation}
    
    If one is to assign a covariant formulation for the interaction of quantum systems with gravity, one must be able to prescribe an energy momentum tensor to the system. If one has a Lagrangian description, there is a natural way to obtain the stress-energy tensor and the coupling with gravity. To do that, assume that the system can be described by a wave function $\psi(\bm x)$  under the influence of a potential $V(\bm x)$ that evolves according to Scr\"odinger's equation. Then, the dynamics can be derived from the following Lagrangian density:
    \begin{equation}
        \begin{gathered}
        \mathcal{L} = \frac{\ii}{2}\psi^*(\bm x) \partial_t \psi(\bm x) - \frac{\ii}{2}\psi(\bm x) \partial_t \psi^*(\bm x) \\ - \frac{1}{2m} \abs{\nabla\psi(\bm x)}^2 - V(\bm x)\abs{\psi(\bm x)}^2.
        \end{gathered}
    \end{equation}
    Indeed, one can easily check that the variation of the above with respect to $\psi^*(\bm x)$ yields Schr\"odinger equation for $\psi(\bm x)$. Furthermore, the conjugate momenta to $\psi(\bm x)$ and $\psi^*(\bm x)$ read
    \begin{align}
        \pi(\bm x) &= \pdv{\mathcal{L}}{(\partial_t\psi(\bm x))} = \frac{\ii}{2} \psi^*(\bm x),\\
        \pi^*(\bm x) &= \pdv{\mathcal{L}}{(\partial_t\psi^*(\bm x))} = -\frac{\ii}{2} \psi(\bm x).
    \end{align}
    From these, we can calculate the Hamiltonian density for the system, which is given by
    \begin{equation}
        \mathcal{H} = \frac{1}{2m}\abs{\nabla\psi(\bm x)}^2 + V(x)\abs{\psi(\bm x)}^2.
    \end{equation}
    This agrees with the well known result from non relativistic quantum mechanics. Even more, integrating by parts, we obtain the Hamiltonian density in terms of the Hamiltonian of the system:
    \begin{equation}
        \bra{\psi}\hat{H}\ket{\psi} = \int \dd^3 \bm x \braket{\psi}{\bm x}\bra{\bm x}\hat{H}\ket{\psi} = \int \dd^3 \bm x \psi^*(\bm x) \hat{\mathcal{H}}\psi(\bm x),
    \end{equation}
    where the operator $\hat{\mathcal{H}}$ is the differential operator that acts on wave functions,
    \begin{equation}
        \hat{\mathcal{H}} = -\frac{1}{2m}\nabla^2 + V(\bm x).
    \end{equation}
    
    Our goal here is to obtain a coupling with gravity in the weak gravitational regime. In this situation, all we have to do is to shift the metric according to $\eta_{\mu\nu} \longmapsto \eta_{\mu\nu} + h_{\mu\nu}$. In this case, because of the gauge in which we got the solution for $h_{\mu\nu}$ in section \ref{lineargrav}, we only need to change the spatial part that shows up in the gradient term, 
    \begin{equation}
        \frac{1}{2m}\abs{\nabla\psi(\bm x)}^2 =\frac{1}{2m} \delta^{ij}\nabla_i\psi^*(\bm x) \nabla_j \psi(\bm x).
    \end{equation}
    The change in the metric due to first order perturbations amounts to ${\delta^{ij} \longmapsto \delta^{ij}+h^{ij}}$. This yields the interaction Hamiltonian
    \begin{equation}
        \frac{1}{2m} \nabla_i\psi^*(\bm x)\nabla_j\psi(\bm x) h^{ij}(t,\bm x).
    \end{equation}
    This can be rewritten in terms of operators in the Hilbert space that we use in non relativistic quantum mechanics. We can integrate by parts to obtain the interaction Hamiltonian in terms of the position and momentum operators in the Hilbert space,
    \begin{equation}\label{HI}
        \hat{H}_I = \frac{1}{2m} h^{ij}(t,\hat{\bm x})\hat{p}_i \hat{p}_j,
    \end{equation}
    in the Schr\"odinger picture.
    This operator can be further rewritten in the fashion of what is done in quantum optics, using the eigenfunctions of the free Hamiltonian. Let $\ket{n}$ be its $n$-th eigenstate, with associated eigenfunction in momentum space given by $\tilde{\psi}_n(\bm p) = \bra{\bm p}\ket{n}$. We can then insert identities in expression \eqref{HI} to obtain
    
    \begin{align}
        &\hat{H}_I\!\! =\! \frac{1}{2m}\!\sum_{nm}\!\! \int\!\! \dd^3\! \bm p \!\!\!\int\!\! \dd^3\!\bm p'\!\!\braket{n}{\bm p}\!\!\bra{\bm p} \!h^{ij}(t,\hat{\bm x}) \hat{p}_i\hat{p}_j\! \ket{\bm p'}\!\!\bra{\bm p'}\ket{m}\!\ket{n}\!\!\bra{m}\nonumber\\
        &=\!\frac{1}{2m}\!\!\sum_{nm} \!\!\int\!\! \dd^3\!\bm p\! \!\!\int\!\! \dd^3\!\bm p'\!\tilde{\psi}_n^*(\bm p)\tilde{\psi}_m(\bm p')p'_ip'_j\! \bra{\bm p}\!h^{ij}(t,\hat{\bm x})\! \ket{\bm p'}\!\ket{n}\!\!\bra{m}\!.\label{HIUgly}
    \end{align}

    Notice that the perturbation of the gravitational field is now evaluated at the position operator $\hat{\bm x}$. We can remove the dependence on the operator and factor out the classical field evaluated at a given position by simply introducing the identity in terms of the position eigenstates,
    \begin{align}
        \bra{\bm p} h^{ij}(t,\hat{\bm x})\ket{\bm p'} &= \bra{\bm p} \left(\int\!\! \dd^3\bm x \ket{\bm x}\bra{\bm x}h^{ij}(t,\hat{\bm x})\right)\ket{\bm p'}\\ 
        &= \frac{1}{(2\pi)^3} \int \!\!\dd^3\bm x h^{ij}(t,\bm x) e^{\ii(\bm p-\bm p')\cdot \bm x}.
    \end{align}
    
    Plugging this back into \eqref{HIUgly}, we get the simplified expression in terms of the real space eigenfunctions $\psi_n(\bm x) = \bra{\bm x}\ket{n}$,
    \begin{equation}
        \hat{H}_I = \frac{1}{2m} \sum_{nm} \int\!\!\dd^3 x  \:\psi_n^*(\bm x)\nabla_i\nabla_j\psi_m(\bm x) h^{ij}(t,\bm x)\ket{n}\!\!\bra{m}.
    \end{equation}
    
    It is then useful to define the smearing tensors for the interaction to be the integrand that couples to $h_{ij}(t,\bm x)$ for each energy level transition, namely,
    
    \begin{equation}
        F^{nm}_{ij}(\bm x) = \nabla_{i}\psi_{n}^*(\bm x) \nabla_{j}\psi_m(\bm x).
    \end{equation}
    Notice the similarity with equation \eqref{fsmearingeg} for the electromagnetic dipole. 
    We therefore obtain the interaction Hamiltonian  in the Schr\"odinger picture to be
    \begin{equation}
         \hat{H}_I = \frac{1}{2m} \sum_{nm} \int\!\!\dd^3 \bm x F^{nm}_{ij}(\bm x) h^{ij}(t,\bm x)\ket{n}\!\!\bra{m}.
    \end{equation}
    
    If we are interested in a specific energy level transition, from a given $n$ to $n+1$, it is useful to relabel the states $n = g$ and $n+1 = e$, associated to a ground and excited states, respectively.  This allows us to write an effective interaction Hamiltonian associated to energy level transitions mediated by interaction with the gravitational field perturbation $h^{ij}(\mathsf x) = h^{ij}(t,\bm x)$,
    \begin{equation}\label{HIfinal}
        \hat{H}_I = \int \dd^3\bm x\left( F_{ij}(\bm x) h^{ij}(\mathsf{x}) \ket{e}\bra{g}+F_{ij}^*(\bm x) h^{ij}(\mathsf x) \ket{g}\bra{e}\right).
    \end{equation}
    Notice that this procedure so far considers the gravitational perturbation to be first quantized, just like the light-matter interaction derivation from \cite{Pozas2016}. In the same spirit as it is done in that case, we then consider the free quantum field expansion from \eqref{modeExpansionPerturbation}. This yields the interaction Hamiltonian for the system and quantum field in the interaction picture to be
    \begin{equation}\label{HIIIII}
        \hat{H}_I(t) = \int \dd^3\bm x\left( F_{ij}(\bm x)e^{\ii\Omega t} \hat{h}^{ij}(\mathsf{x}) \ket{e}\bra{g}+\mathrm{H.c.}\right),
    \end{equation}
    where $\Omega$ is the energy gap between the levels labeled by $e$ and $g$. The interaction Hamiltonian \eqref{HIIIII} is very similar to the one obtained in electromagnetism for the light-matter interaction in atoms from equation \eqref{HIEM}.

    In the electromagnetic case, if one disregards exchanges of angular momentum between detectors and field, one can take the Unruh-DeWitt model as an approximation to describe the interaction. Therefore the scalar approximation can still be able to grasp the fundamental features of entanglement harvesting in the gravitational setting, even though the field $\hat{h}^{ij}(\mathsf{x})$ has spin two. Moreover, it has also been shown that if no scalar approximation is employed in the light-matter interaction model, the entanglement harvesting results can even be enhanced \cite{Pozas2016}.

\section{The Setup}\label{Discussions}

Although it could be seen as natural to expect the fundamental description of nature not to be classical, up to the present day we have no experimental evidence that the gravitational field is quantum. We do have many physical theories for an effective quantum theory of gravity, but none of them has been experimentally tested. In this line of thought, Bose, Marletto and Vedral have proposed an experiment (the BMV experiment) that would be able to  test for the existence of a quantum degree of freedom for the field. However, in their treatment a quantum field theory of gravity is never explicitly considered. 

From the tools developed in section \ref{sectionCoupling}, we see that the weak gravity interaction Hamiltonian is similar to the light-matter interaction one, usually used in quantum optics. We do know that the latter is well approximated by the model of Unruh-DeWitt detectors, in which we set the coupling constant to be the light-matter one. We thus have no reason to expect differently for the gravity-matter interaction in the regime of weak gravitational fields. Therefore, we can apply the techniques of Unruh-DeWitt detectors to describe such interactions.

The setup proposed here is to consider two non relativistic quantum systems, such as atoms, that interact with independent vacua of the electromagnetic field. This can be done by separating the atoms in two different metallic cavities. These are known to act as a boundary conditions that separate the vacuum of the electromagnetic field. Then, the only field that would be able to communicate between the two systems would be the gravitational one. In this setup we then perform the entanglement harvesting procedure between the two detectors.

If any entanglement could be measured between the systems, we would therefore conclude that it must have come from the interaction with the vacuum of the gravitational field. From that we would be able to infer that the gravitational field is indeed a quantum field on a fundamental level, for according to subsection \ref{sectionNoHarvest}, no classical field would be able to produce entanglement between the two systems.

\begin{figure}
    \centering
    \includegraphics[scale = 0.45]{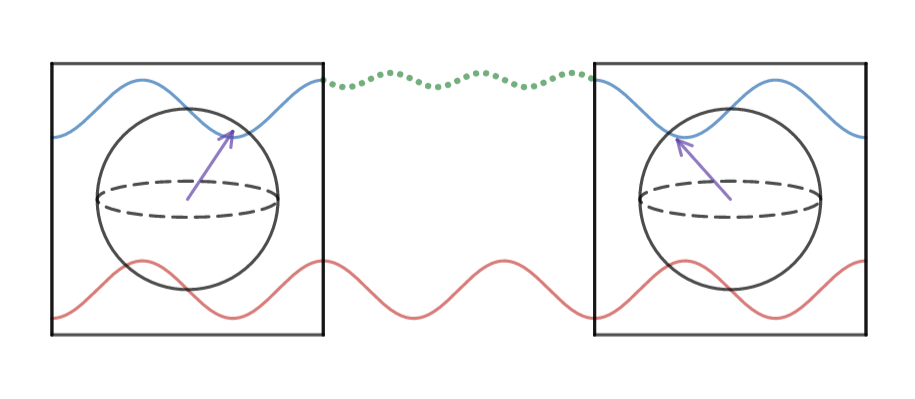}
    \caption{A schematic view of the setup. The electromagnetic field inside the cavities (blue) is independent of the one outside (green). This prevents the atoms to get entangled via the electromagnetic vacuum, and only the gravitational field (red) penetrates both cavities.}
\end{figure}

We can estimate the orders of magnitude related to a real experiment, comparing to the precision we would need for harvesting entanglement from the electromagnetic vacuum only. If one works with systems of the size of atoms, the relevant length scale is of the order of the Bohr radius, which is $a_0 \approx 10^{24}\: \ell_p$, where $\ell_p$ is the Planck length. 
We have seen that the coupling with the gravitational perturbations goes as
\begin{equation}
    \frac{1}{2m}p_ip_j h^{ij},
\end{equation}
while the electromagnetic one is given by
\begin{equation}
    \frac{1}{m} p_i A^i. 
\end{equation}
Taking into consideration that the vacuum fluctuations of the quantum fields themselves depend only on the quantization procedure, the order of magnitude of $\textrm{M}_p h^{ij}$ and $A^i$ in the vacuum is the same. We then find the ration between the expectation value of the two interaction terms to be
\begin{equation}
    \frac{\ev{\frac{1}{2m}p_ip_j h^{ij}}}{\ev{\frac{1}{m}p_iA^i}} \propto \frac{\ev{p}}{\textrm{M}_p},
\end{equation}
where we recall that $\ev{p}$ should also be measured in Planck units. For an atom, we obtain that $\ev{p} \approx 1/a_0 \approx 10^{-24}\:\textrm{M}_p$. From that, we get that the gravitational effects to be observed are $24$ orders of magnitude less than the electromagnetic counterpart.

This means that our setup is not feasible with current technology. Indeed, as of today, we have no experimental results on entanglement harvesting. 
Moreover, the experimental challenges range from the precision of the necessary quantum state tomography to the reliability of the shielding of the cavity. 
One may then wonder how it would be possible to try to partially circumvent some of those challenges; in order to do that, some specific features of our model can be exploited. One of them is the fact that the gravitational perturbation is a spin-2 field, whereas the electromagnetic field has spin-1. Therefore, the gravitational quadrupole coupling allows for atomic transitions that are not possible with the electromagnetic dipole coupling. 
The challenge of reducing electromagnetic-induced entanglement could also be solved by other means than cavities. For instance, some work has been done on trying to prevent entanglement between timelike-separated atoms using an intermediate system \cite{neweduardo}, which could also be a strategy to further shield  electromagnetically induced entanglement in this scenario.

In summary, however, if it would be possible to perform an experiment associated to this setup, positive results would provide strong evidence for the quantum nature of the gravitational field.

\section{Conclusions}

    We have presented a detector-based framework that could probe the quantum nature of gravity by coupling two quantum systems to a gravitational field. If the field is indeed quantum, the systems should interact with it and become entangled between themselves. This setup is, therefore, a promising way to probe quantum gravitational effects in the linearized gravity regime.
    
    Furthermore, we have developed a method for treating the interaction of gravitational perturbations with matter using a Lagrangian for the wavefunction of the system. This method would also be able to reproduce the electromagnetic interactions from \cite{Pozas2016} by using the standard non relativistic approach for electromagnetism with continuous quantum systems.
    
    The model presented here can be compared to the BMV experiment for, just like it, we propose a setup to test the fundamental nature of gravity. In the BMV model, quantum particles are considered to be sources of the gravitational field and it is treated in the Newtonian approximation. It is then not obvious that such a behaviour would imply a quantum nature for the gravitational field, as was mentioned in \cite{BeiLok}. It should also be pointed out that the non-relativistic approach may not contemplate the features of a full quantum theory of gravity. 
    On the other hand, in the model proposed in this article, we compare the treatment of the gravitational field as a quantum field with its classical analogue and harvest entanglement from the vacuum itself to obtain a conclusion. 
    When comparing to the BMV experiment, our proposal is by no means a relativistic treatment of the same idea, but a complementary way of considering the problem of measuring quantum effects of the gravitational field, motivated by entanglement harvesting techniques.
    
    Although the experimental feasibility of the model proposed here cannot be expected to be achieved in the next years, it opens the way to new ideas and has the advantage of not requiring high energy scales, only low energy detectors coupled to the vacuum. Furthermore, we have provided a formalism that justifies the uses of UDW detector models for probing linearized quantum gravity.
    

\section{Acknowledgements}
    The authors thank Flaminia Giacomini, Anne-Catherine De La Hamette and Jonas Neuser for insightful discussions, as well as Ghislaine Coulter-de Wit, Jonas Neuser, Eduardo Mart\'{i}n-Mart\'{i}nez and Jos\'{e} de Ram\'{o}n for helping review the text. Special thanks are given to Perimeter Institute for Theoretical Physics and for the Perimeter Scholars International program. Research at Perimeter Institute is supported in part by the Government
of Canada through the Department of Innovation, Science and Economic Development Canada and by the
Province of Ontario through the Ministry of Economic
Development, Job Creation and Trade. T.R.P. and B.S.L.T. thank
IFT-UNESP/ICTP-SAIFR and CAPES for partial financial support. 
   R.F. thanks École Normale Supérieure de Lyon for partial financial support.
    
    

\bibliography{references}

\begin{thebibliography}{23}%
\makeatletter
\providecommand \@ifxundefined [1]{%
 \@ifx{#1\undefined}
}%
\providecommand \@ifnum [1]{%
 \ifnum #1\expandafter \@firstoftwo
 \else \expandafter \@secondoftwo
 \fi
}%
\providecommand \@ifx [1]{%
 \ifx #1\expandafter \@firstoftwo
 \else \expandafter \@secondoftwo
 \fi
}%
\providecommand \natexlab [1]{#1}%
\providecommand \enquote  [1]{``#1''}%
\providecommand \bibnamefont  [1]{#1}%
\providecommand \bibfnamefont [1]{#1}%
\providecommand \citenamefont [1]{#1}%
\providecommand \href@noop [0]{\@secondoftwo}%
\providecommand \href [0]{\begingroup \@sanitize@url \@href}%
\providecommand \@href[1]{\@@startlink{#1}\@@href}%
\providecommand \@@href[1]{\endgroup#1\@@endlink}%
\providecommand \@sanitize@url [0]{\catcode `\\12\catcode `\$12\catcode
  `\&12\catcode `\#12\catcode `\^12\catcode `\_12\catcode `\%12\relax}%
\providecommand \@@startlink[1]{}%
\providecommand \@@endlink[0]{}%
\providecommand \url  [0]{\begingroup\@sanitize@url \@url }%
\providecommand \@url [1]{\endgroup\@href {#1}{\urlprefix }}%
\providecommand \urlprefix  [0]{URL }%
\providecommand \Eprint [0]{\href }%
\providecommand \doibase [0]{http://dx.doi.org/}%
\providecommand \selectlanguage [0]{\@gobble}%
\providecommand \bibinfo  [0]{\@secondoftwo}%
\providecommand \bibfield  [0]{\@secondoftwo}%
\providecommand \translation [1]{[#1]}%
\providecommand \BibitemOpen [0]{}%
\providecommand \bibitemStop [0]{}%
\providecommand \bibitemNoStop [0]{.\EOS\space}%
\providecommand \EOS [0]{\spacefactor3000\relax}%
\providecommand \BibitemShut  [1]{\csname bibitem#1\endcsname}%
\let\auto@bib@innerbib\@empty
\bibitem [{\citenamefont {Marletto}\ and\ \citenamefont {Vedral}(2017)}]{BMV1}%
  \BibitemOpen
  \bibfield  {author} {\bibinfo {author} {\bibfnamefont {C.}~\bibnamefont
  {Marletto}}\ and\ \bibinfo {author} {\bibfnamefont {V.}~\bibnamefont
  {Vedral}},\ }\href {\doibase 10.1103/PhysRevLett.119.240402} {\bibfield
  {journal} {\bibinfo  {journal} {Phys. Rev. Lett.}\ }\textbf {\bibinfo
  {volume} {119}},\ \bibinfo {pages} {240402} (\bibinfo {year}
  {2017})}\BibitemShut {NoStop}%
\bibitem [{\citenamefont {Bose}\ \emph {et~al.}(2017)\citenamefont {Bose},
  \citenamefont {Mazumdar}, \citenamefont {Morley}, \citenamefont {Ulbricht},
  \citenamefont {Toro\ifmmode~\check{s}\else \v{s}\fi{}}, \citenamefont
  {Paternostro}, \citenamefont {Geraci}, \citenamefont {Barker}, \citenamefont
  {Kim},\ and\ \citenamefont {Milburn}}]{BMV2}%
  \BibitemOpen
  \bibfield  {author} {\bibinfo {author} {\bibfnamefont {S.}~\bibnamefont
  {Bose}}, \bibinfo {author} {\bibfnamefont {A.}~\bibnamefont {Mazumdar}},
  \bibinfo {author} {\bibfnamefont {G.~W.}\ \bibnamefont {Morley}}, \bibinfo
  {author} {\bibfnamefont {H.}~\bibnamefont {Ulbricht}}, \bibinfo {author}
  {\bibfnamefont {M.}~\bibnamefont {Toro\ifmmode~\check{s}\else \v{s}\fi{}}},
  \bibinfo {author} {\bibfnamefont {M.}~\bibnamefont {Paternostro}}, \bibinfo
  {author} {\bibfnamefont {A.~A.}\ \bibnamefont {Geraci}}, \bibinfo {author}
  {\bibfnamefont {P.~F.}\ \bibnamefont {Barker}}, \bibinfo {author}
  {\bibfnamefont {M.~S.}\ \bibnamefont {Kim}}, \ and\ \bibinfo {author}
  {\bibfnamefont {G.}~\bibnamefont {Milburn}},\ }\href {\doibase
  10.1103/PhysRevLett.119.240401} {\bibfield  {journal} {\bibinfo  {journal}
  {Phys. Rev. Lett.}\ }\textbf {\bibinfo {volume} {119}},\ \bibinfo {pages}
  {240401} (\bibinfo {year} {2017})}\BibitemShut {NoStop}%
\bibitem [{\citenamefont {Anastopoulos}\ and\ \citenamefont
  {Hu}(2018)}]{BeiLok}%
  \BibitemOpen
  \bibfield  {author} {\bibinfo {author} {\bibfnamefont {C.}~\bibnamefont
  {Anastopoulos}}\ and\ \bibinfo {author} {\bibfnamefont {B.-L.}\ \bibnamefont
  {Hu}},\ }\href@noop {} {\enquote {\bibinfo {title} {Comment on ``a spin
  entanglement witness for quantum gravity'' and on ``gravitationally induced
  entanglement between two massive particles is sufficient evidence of quantum
  effects in gravity"},}\ } (\bibinfo {year} {2018}),\ \Eprint
  {http://arxiv.org/abs/1804.11315} {arXiv:1804.11315 [quant-ph]} \BibitemShut
  {NoStop}%
\bibitem [{\citenamefont {Anastopoulos}\ and\ \citenamefont
  {Hu}(2015)}]{BeiLok2}%
  \BibitemOpen
  \bibfield  {author} {\bibinfo {author} {\bibfnamefont {C.}~\bibnamefont
  {Anastopoulos}}\ and\ \bibinfo {author} {\bibfnamefont {B.~L.}\ \bibnamefont
  {Hu}},\ }\href {\doibase 10.1088/0264-9381/32/16/165022} {\bibfield
  {journal} {\bibinfo  {journal} {Classical and Quantum Gravity}\ }\textbf
  {\bibinfo {volume} {32}},\ \bibinfo {pages} {165022} (\bibinfo {year}
  {2015})}\BibitemShut {NoStop}%
\bibitem [{\citenamefont {Belenchia}\ \emph {et~al.}(2018)\citenamefont
  {Belenchia}, \citenamefont {Wald}, \citenamefont {Giacomini}, \citenamefont
  {Castro-Ruiz}, \citenamefont {Brukner},\ and\ \citenamefont
  {Aspelmeyer}}]{flaminia}%
  \BibitemOpen
  \bibfield  {author} {\bibinfo {author} {\bibfnamefont {A.}~\bibnamefont
  {Belenchia}}, \bibinfo {author} {\bibfnamefont {R.~M.}\ \bibnamefont {Wald}},
  \bibinfo {author} {\bibfnamefont {F.}~\bibnamefont {Giacomini}}, \bibinfo
  {author} {\bibfnamefont {E.}~\bibnamefont {Castro-Ruiz}}, \bibinfo {author}
  {\bibfnamefont {i.~c.~v.}\ \bibnamefont {Brukner}}, \ and\ \bibinfo {author}
  {\bibfnamefont {M.}~\bibnamefont {Aspelmeyer}},\ }\href {\doibase
  10.1103/PhysRevD.98.126009} {\bibfield  {journal} {\bibinfo  {journal} {Phys.
  Rev. D}\ }\textbf {\bibinfo {volume} {98}},\ \bibinfo {pages} {126009}
  (\bibinfo {year} {2018})}\BibitemShut {NoStop}%
\bibitem [{\citenamefont {Marshman}\ \emph {et~al.}(2019)\citenamefont
  {Marshman}, \citenamefont {Mazumdar},\ and\ \citenamefont {Bose}}]{Mazumdar}%
  \BibitemOpen
  \bibfield  {author} {\bibinfo {author} {\bibfnamefont {R.~J.}\ \bibnamefont
  {Marshman}}, \bibinfo {author} {\bibfnamefont {A.}~\bibnamefont {Mazumdar}},
  \ and\ \bibinfo {author} {\bibfnamefont {S.}~\bibnamefont {Bose}},\
  }\href@noop {} {\enquote {\bibinfo {title} {Locality \& entanglement in
  table-top testing of the quantum nature of linearized gravity},}\ } (\bibinfo
  {year} {2019}),\ \Eprint {http://arxiv.org/abs/1907.01568} {arXiv:1907.01568
  [quant-ph]} \BibitemShut {NoStop}%
\bibitem [{\citenamefont {Weinberg}(1965)}]{weinberg}%
  \BibitemOpen
  \bibfield  {author} {\bibinfo {author} {\bibfnamefont {S.}~\bibnamefont
  {Weinberg}},\ }\href {\doibase 10.1103/PhysRev.138.B988} {\bibfield
  {journal} {\bibinfo  {journal} {Phys. Rev.}\ }\textbf {\bibinfo {volume}
  {138}},\ \bibinfo {pages} {B988} (\bibinfo {year} {1965})}\BibitemShut
  {NoStop}%
\bibitem [{\citenamefont {Carroll}(2019)}]{carroll}%
  \BibitemOpen
  \bibfield  {author} {\bibinfo {author} {\bibfnamefont {S.~M.}\ \bibnamefont
  {Carroll}},\ }\href {\doibase 10.1017/9781108770385} {\emph {\bibinfo {title}
  {Spacetime and Geometry: An Introduction to General Relativity}}}\ (\bibinfo
  {publisher} {Cambridge University Press},\ \bibinfo {year}
  {2019})\BibitemShut {NoStop}%
\bibitem [{\citenamefont {DeWitt}(1980)}]{DeWitt}%
  \BibitemOpen
  \bibfield  {author} {\bibinfo {author} {\bibfnamefont {B.}~\bibnamefont
  {DeWitt}},\ }\href@noop {} {\emph {\bibinfo {title} {General Relativity; an
  Einstein Centenary Survey}}}\ (\bibinfo  {publisher} {Cambridge University
  Press},\ \bibinfo {address} {Cambridge, UK},\ \bibinfo {year}
  {1980})\BibitemShut {NoStop}%
\bibitem [{\citenamefont {Unruh}\ and\ \citenamefont
  {Wald}(1984)}]{Unruh-Wald}%
  \BibitemOpen
  \bibfield  {author} {\bibinfo {author} {\bibfnamefont {W.~G.}\ \bibnamefont
  {Unruh}}\ and\ \bibinfo {author} {\bibfnamefont {R.~M.}\ \bibnamefont
  {Wald}},\ }\href {\doibase 10.1103/PhysRevD.29.1047} {\bibfield  {journal}
  {\bibinfo  {journal} {Phys. Rev. D}\ }\textbf {\bibinfo {volume} {29}},\
  \bibinfo {pages} {1047} (\bibinfo {year} {1984})}\BibitemShut {NoStop}%
\bibitem [{\citenamefont {Valentini}(1991)}]{Valentini1991}%
  \BibitemOpen
  \bibfield  {author} {\bibinfo {author} {\bibfnamefont {A.}~\bibnamefont
  {Valentini}},\ }\href {\doibase
  http://dx.doi.org/10.1016/0375-9601(91)90952-5} {\bibfield  {journal}
  {\bibinfo  {journal} {Phys. Lett. A}\ }\textbf {\bibinfo {volume} {153}},\
  \bibinfo {pages} {321 } (\bibinfo {year} {1991})}\BibitemShut {NoStop}%
\bibitem [{\citenamefont {Reznik}\ \emph {et~al.}(2005)\citenamefont {Reznik},
  \citenamefont {Retzker},\ and\ \citenamefont {Silman}}]{Reznik1}%
  \BibitemOpen
  \bibfield  {author} {\bibinfo {author} {\bibfnamefont {B.}~\bibnamefont
  {Reznik}}, \bibinfo {author} {\bibfnamefont {A.}~\bibnamefont {Retzker}}, \
  and\ \bibinfo {author} {\bibfnamefont {J.}~\bibnamefont {Silman}},\ }\href
  {http://link.aps.org/abstract/PRA/v71/e042104} {\bibfield  {journal}
  {\bibinfo  {journal} {Phys. Rev. A}\ }\textbf {\bibinfo {volume} {71}},\
  \bibinfo {eid} {042104} (\bibinfo {year} {2005})}\BibitemShut {NoStop}%
\bibitem [{\citenamefont {Pozas-Kerstjens}\ and\ \citenamefont
  {Mart\'{i}n-Mart\'{i}nez}(2015)}]{Pozas-Kerstjens:2015}%
  \BibitemOpen
  \bibfield  {author} {\bibinfo {author} {\bibfnamefont {A.}~\bibnamefont
  {Pozas-Kerstjens}}\ and\ \bibinfo {author} {\bibfnamefont {E.}~\bibnamefont
  {Mart\'{i}n-Mart\'{i}nez}},\ }\href {\doibase 10.1103/PhysRevD.92.064042}
  {\bibfield  {journal} {\bibinfo  {journal} {Phys. Rev. D}\ }\textbf {\bibinfo
  {volume} {92}},\ \bibinfo {pages} {064042} (\bibinfo {year}
  {2015})}\BibitemShut {NoStop}%
\bibitem [{\citenamefont {Ng}\ \emph {et~al.}(2018)\citenamefont {Ng},
  \citenamefont {Mann},\ and\ \citenamefont {Martín-Martínez}}]{robMann1}%
  \BibitemOpen
  \bibfield  {author} {\bibinfo {author} {\bibfnamefont {K.~K.}\ \bibnamefont
  {Ng}}, \bibinfo {author} {\bibfnamefont {R.~B.}\ \bibnamefont {Mann}}, \ and\
  \bibinfo {author} {\bibfnamefont {E.}~\bibnamefont {Martín-Martínez}},\
  }\href {\doibase 10.1103/PhysRevD.98.125005} {\bibfield  {journal} {\bibinfo
  {journal} {Phys. Rev. D}\ }\textbf {\bibinfo {volume} {98}},\ \bibinfo
  {pages} {125005} (\bibinfo {year} {2018})},\ \Eprint
  {http://arxiv.org/abs/1809.06878} {arXiv:1809.06878 [quant-ph]} \BibitemShut
  {NoStop}%
\bibitem [{\citenamefont {Henderson}\ \emph {et~al.}(2018)\citenamefont
  {Henderson}, \citenamefont {Hennigar}, \citenamefont {Mann}, \citenamefont
  {Smith},\ and\ \citenamefont {Zhang}}]{robMann2}%
  \BibitemOpen
  \bibfield  {author} {\bibinfo {author} {\bibfnamefont {L.~J.}\ \bibnamefont
  {Henderson}}, \bibinfo {author} {\bibfnamefont {R.~A.}\ \bibnamefont
  {Hennigar}}, \bibinfo {author} {\bibfnamefont {R.~B.}\ \bibnamefont {Mann}},
  \bibinfo {author} {\bibfnamefont {A.~R.}\ \bibnamefont {Smith}}, \ and\
  \bibinfo {author} {\bibfnamefont {J.}~\bibnamefont {Zhang}},\ }\href
  {\doibase 10.1088/1361-6382/aae27e} {\bibfield  {journal} {\bibinfo
  {journal} {Class. Quant. Grav.}\ }\textbf {\bibinfo {volume} {35}},\ \bibinfo
  {pages} {21LT02} (\bibinfo {year} {2018})},\ \Eprint
  {http://arxiv.org/abs/1712.10018} {arXiv:1712.10018 [quant-ph]} \BibitemShut
  {NoStop}%
\bibitem [{\citenamefont {Pozas-Kerstjens}\ and\ \citenamefont
  {Mart\'{i}n-Mart\'{i}nez}(2016)}]{Pozas2016}%
  \BibitemOpen
  \bibfield  {author} {\bibinfo {author} {\bibfnamefont {A.}~\bibnamefont
  {Pozas-Kerstjens}}\ and\ \bibinfo {author} {\bibfnamefont {E.}~\bibnamefont
  {Mart\'{i}n-Mart\'{i}nez}},\ }\href {\doibase 10.1103/PhysRevD.94.064074}
  {\bibfield  {journal} {\bibinfo  {journal} {Phys. Rev. D}\ }\textbf {\bibinfo
  {volume} {94}},\ \bibinfo {pages} {064074} (\bibinfo {year}
  {2016})}\BibitemShut {NoStop}%
\bibitem [{\citenamefont {Mart\'{i}n-Mart\'{i}nez}\ and\ \citenamefont
  {Rodriguez-Lopez}(2018)}]{eduardo}%
  \BibitemOpen
  \bibfield  {author} {\bibinfo {author} {\bibfnamefont {E.}~\bibnamefont
  {Mart\'{i}n-Mart\'{i}nez}}\ and\ \bibinfo {author} {\bibfnamefont
  {P.}~\bibnamefont {Rodriguez-Lopez}},\ }\href {\doibase
  10.1103/PhysRevD.97.105026} {\bibfield  {journal} {\bibinfo  {journal} {Phys.
  Rev. D}\ }\textbf {\bibinfo {volume} {97}},\ \bibinfo {pages} {105026}
  (\bibinfo {year} {2018})}\BibitemShut {NoStop}%
\bibitem [{\citenamefont {Mart\'{\i}n-Mart\'{\i}nez}\ \emph
  {et~al.}(2020)\citenamefont {Mart\'{\i}n-Mart\'{\i}nez}, \citenamefont
  {Perche},\ and\ \citenamefont {de~S.~L.~Torres}}]{us}%
  \BibitemOpen
  \bibfield  {author} {\bibinfo {author} {\bibfnamefont {E.}~\bibnamefont
  {Mart\'{\i}n-Mart\'{\i}nez}}, \bibinfo {author} {\bibfnamefont {T.~R.}\
  \bibnamefont {Perche}}, \ and\ \bibinfo {author} {\bibfnamefont
  {B.}~\bibnamefont {de~S.~L.~Torres}},\ }\href {\doibase
  10.1103/PhysRevD.101.045017} {\bibfield  {journal} {\bibinfo  {journal}
  {Phys. Rev. D}\ }\textbf {\bibinfo {volume} {101}},\ \bibinfo {pages}
  {045017} (\bibinfo {year} {2020})}\BibitemShut {NoStop}%
\bibitem [{\citenamefont {Unruh}(1976)}]{Unruh1976}%
  \BibitemOpen
  \bibfield  {author} {\bibinfo {author} {\bibfnamefont {W.~G.}\ \bibnamefont
  {Unruh}},\ }\href {\doibase 10.1103/PhysRevD.14.870} {\bibfield  {journal}
  {\bibinfo  {journal} {Phys. Rev. D}\ }\textbf {\bibinfo {volume} {14}},\
  \bibinfo {pages} {870} (\bibinfo {year} {1976})}\BibitemShut {NoStop}%
\bibitem [{\citenamefont {Candelas}\ and\ \citenamefont
  {Sciama}(1977)}]{Sciama1977}%
  \BibitemOpen
  \bibfield  {author} {\bibinfo {author} {\bibfnamefont {P.}~\bibnamefont
  {Candelas}}\ and\ \bibinfo {author} {\bibfnamefont {D.~W.}\ \bibnamefont
  {Sciama}},\ }\href {\doibase 10.1103/PhysRevLett.38.1372} {\bibfield
  {journal} {\bibinfo  {journal} {Phys. Rev. Lett.}\ }\textbf {\bibinfo
  {volume} {38}},\ \bibinfo {pages} {1372} (\bibinfo {year}
  {1977})}\BibitemShut {NoStop}%
\bibitem [{\citenamefont {Hotta}(2009)}]{teleportation}%
  \BibitemOpen
  \bibfield  {author} {\bibinfo {author} {\bibfnamefont {M.}~\bibnamefont
  {Hotta}},\ }\href {\doibase 10.1143/JPSJ.78.034001} {\bibfield  {journal}
  {\bibinfo  {journal} {Journal of the Physical Society of Japan}\ }\textbf
  {\bibinfo {volume} {78}},\ \bibinfo {pages} {034001} (\bibinfo {year}
  {2009})},\ \Eprint
  {http://arxiv.org/abs/https://doi.org/10.1143/JPSJ.78.034001}
  {https://doi.org/10.1143/JPSJ.78.034001} \BibitemShut {NoStop}%
\bibitem [{\citenamefont {Silman}\ and\ \citenamefont
  {Reznik}(2007)}]{reznik2}%
  \BibitemOpen
  \bibfield  {author} {\bibinfo {author} {\bibfnamefont {J.}~\bibnamefont
  {Silman}}\ and\ \bibinfo {author} {\bibfnamefont {B.}~\bibnamefont
  {Reznik}},\ }\href@noop {} {\bibfield  {journal} {\bibinfo  {journal} {Phys.
  Rev. A}\ }\textbf {\bibinfo {volume} {75}},\ \bibinfo {pages} {052307}
  (\bibinfo {year} {2007})}\BibitemShut {NoStop}%
\bibitem [{\citenamefont {Sahu}\ \emph {et~al.}(tion)\citenamefont {Sahu},
  \citenamefont {Melgarejo-Lermas},\ and\ \citenamefont
  {Mart\'{i}n-Mart\'{i}nez}}]{neweduardo}%
  \BibitemOpen
  \bibfield  {author} {\bibinfo {author} {\bibfnamefont {A.}~\bibnamefont
  {Sahu}}, \bibinfo {author} {\bibfnamefont {I.}~\bibnamefont
  {Melgarejo-Lermas}}, \ and\ \bibinfo {author} {\bibfnamefont
  {E.}~\bibnamefont {Mart\'{i}n-Mart\'{i}nez}},\ }\href@noop {} {\enquote
  {\bibinfo {title} {Cancelling the harvesting of correlations in {QFT}},}\ }
  (\bibinfo {year} {In preparation})\BibitemShut {NoStop}%
\end{thebibliography}%

\end{document}